\documentclass[fleqn,12pt,twoside]{article}
\usepackage{espcrc1}

\usepackage{graphicx}
\usepackage[figuresright]{rotating}

\newcommand{\MeV}{\rm MeV}
\newcommand{\GeV}{\rm GeV}

\newcommand{\AmS}{{\protect\the\textfont2
  A\kern-.1667em\lower.5ex\hbox{M}\kern-.125emS}}

\title{Polarized Parton Distributions and QCD Analysis}

\author{J. Bl\"umlein and H. B\"ottcher\\
        Deutsches Elektronen Synchrotron DESY, Platanenallee 6,
        D--15738 Zeuthen, Germany}%
       
\begin{document}
\sloppy

\maketitle

\begin{abstract}

\noindent
\small
In this talk we summarize main results of a recent determination of the
polarized deeply inelastic parton distributions to NLO from the world
data. In the analysis the LO and NLO parton densities and their $1\sigma$
statistical errors were derived and parameterized. The strong coupling
constant $\alpha_s(M_Z^2)$ is determined $\alpha_s(M_Z^2) = 0.113 \pm
0.004 {\rm (stat.)} \pm~0.004 {\rm (fac.)} +0.008/-0.005 {\rm (ren.)}$
Comparisons of the low moments of the parton densities with recent
lattice results are given. A detailed error-analysis of the gluon
density is performed.
\normalsize
\end{abstract}

\section{Introduction}

\noindent
The polarized parton densities of deep inelastic scattering can be
unfolded from measurements of the longitudinal polarization asymmetry.
Here the unpolarized denominator--function serves as a normalization and
is usually also being determined from data \cite{a1,a2}. 
The present polarized deep inelastic world data
have reached a precision which allows to extract the corresponding
parton densities and their correlated statistical errors as a function of
$x$ and $Q^2$ as well as the QCD $\Lambda$--parameter. Previous analyses
\cite{a3,a4} were limited to a
determination of the central value of the parton densities. In a recent
analysis \cite{a5} we extended the investigation
using the available world data to determine also the correlated errors 
and the QCD--scale, for which numerical parameterizations are provided
as a {\tt FORTRAN} code. In this analysis the QCD scale has been also
determined using  scheme--invariant evolution equations, which directly
describe the evolution of observables as the structure function 
$g_1(x,Q^2)$ and its slope $\partial g_1(x,Q^2)/\partial \ln(Q^2)$ which
was yielding nearly the same value for $\alpha_s(M_Z^2)$. Finally the
lower Mellin moments of the present parton distributions were calculated
and compared to results obtained in Lattice--QCD. This Note summarizes
main results of our analysis.
\section{The Parameterizations}

\noindent
The principle shape of the polarized parton densities $x\Delta u_v, 
x \Delta d_v,
x\Delta q_{\rm sea}$ and $x\Delta G$ were chosen as
\begin{equation}
x \Delta q_i(x,Q_0^2) = \eta_i A_i x^{a_i} (1-x)^{b_i} (1 + \gamma_i x
+ \rho_i x^{1/2})~,
\end{equation}
with $A_i$ the normalization and $\eta_i$  the first moment. 
All details of the analysis are found in \cite{a5}.
Some parameters are fixed by sum rules. 
It turned out that not all parameters can 
be measured at the same quality and have to be set to their value at
$\chi^2_{min}$. In Fig.~1 we show results on $x\Delta G(x,Q^2)$
emphasizing
the correlated statistical errors and the present experimental and
theoretical systematics. It is evident that a further improvement of
the statistical error very soon will require
3--loop evolution equations 
to cope with the forthcoming experimental errors.
\section{The strong coupling constant}

\noindent
In the analysis $\alpha_s(M_Z^2)$ is measured at NLO. We also include the
variations due to a change in the renormalization and factorization scales
varying $\mu_{R,F}^2$ by a factor of 2 around $Q^2$. We obtain
\begin{equation}
\alpha_s(M_Z^2) = 0.113 \pm 0.004 (stat) \pm 0.004 (fac) +0.008/-0.005 (ren). 
\end{equation}
\section{Comparison with lattice results}

\noindent
The parameterizations may be used to evaluate integer Mellin moments
of the respective parton distributions including their errors. These
moments correspond to the experimental data, as far as the parameters of 
the corresponding representations are fixed in the QCD--fit procedure.
To evaluate moments the parameterizations have to be partly extrapolated
outside the domain of the current measurements. It is now interesting to 
compare these moments to the results which have been found in recent
lattice flavor non--singlet simulations (see Table~1). The comparison 
shows that the extractions do well agree in all the nine values determined
in the lattice measurements~\footnote{The value of $g_A$, the first
moment of $\Delta u_v - \Delta d_v$ seemed to have moved 
by $\sim +10\%$ in a very
recent lattice measurement (Panic'02).}. The observation of this 
agreement,
despite the use of pion masses $m_\pi~\sim~600 \MeV$ in the lattice
simulations, was surprising to us in early 2002, since parabolic 
extrapolations were needed to reach agreement between experimental and
lattice measurements in the unpolarized case. However, later the year it 
was found, that the extrapolation to the chiral limit is indeed nearly 
flat \cite{a6}. The reason for this
is a cancellation due to $\Delta$--baryon contributions.
%
\renewcommand{\arraystretch}{1.2}
%
\begin{center}
\begin{tabular}{|c|r|r|r|r|}
\hline 
\hline 
\multicolumn{1}{|l|}{ }&
\multicolumn{1}{c|}{ }&
\multicolumn{1}{c|}{\tt QCD results }&
\multicolumn{2}{c|}{ lattice results } \\
\cline{3-5}
\multicolumn{1}{|c|}{ $\Delta f$ }& 
\multicolumn{1}{ c|}{ $n$ } &
\multicolumn{1}{ c|}{NLO moments} &
\multicolumn{1}{ c|}{QCDSF    } &
\multicolumn{1}{ c|}{LHPC/    } \\
\multicolumn{1}{|c|}{ }&
\multicolumn{1}{ c|}{   } &
\multicolumn{1}{ c|}{ at $Q^2=4~\rm{GeV^2}$ } &
\multicolumn{1}{ c|}{         } &
\multicolumn{1}{ c|}{SESAM    } \\
\hline \hline
$\Delta u_v$  &--1 & $0.926 \pm 0.071$ & 0.889(29) & 0.860(69) \\
              &  0 & $0.163 \pm 0.014$ & 0.198(8) & 0.242(22) \\
              &  1 & $0.055 \pm 0.006$ & 0.041(9) & 0.116(42) \\
\hline
$\Delta d_v$  &--1 & $-0.341 \pm 0.123$ & -0.236(27) & -0.171(43) \\
              &  0 & $-0.047 \pm 0.021$ & -0.048(3) & -0.029(13) \\
              &  1 & $-0.015 \pm 0.009$ & -0.028(2) &  0.001(25) \\
\hline
$\Delta u$--$\Delta d$
              &--1 & $1.267  \pm 0.142$ &  1.14(3) &  1.031(81) \\
              &  0 & $0.210  \pm 0.025$ &  0.246(9) &  0.271(25) \\
              &  1 & $0.070 \pm 0.011 $ &  0.069(9) &  0.115(49) \\
\hline \hline
\end{tabular}
\end{center}
\renewcommand{\arraystretch}{1}
%
{\footnotesize
 {\bf Table 1.}~~{Moments of the NLO parton densities (for ISET =3, see
\cite{a5}) at $Q^2 = 4~\rm{GeV^2}$ and from
recent lattice simulations at the scale 
$\mu^2 = 1/a^2 \sim 4~\rm{GeV^2}$, see \cite{a7}.}}

\vspace*{-0.15cm}
\begin{figure}[t]
\begin{center}
\includegraphics[angle=0, width=10.4cm]{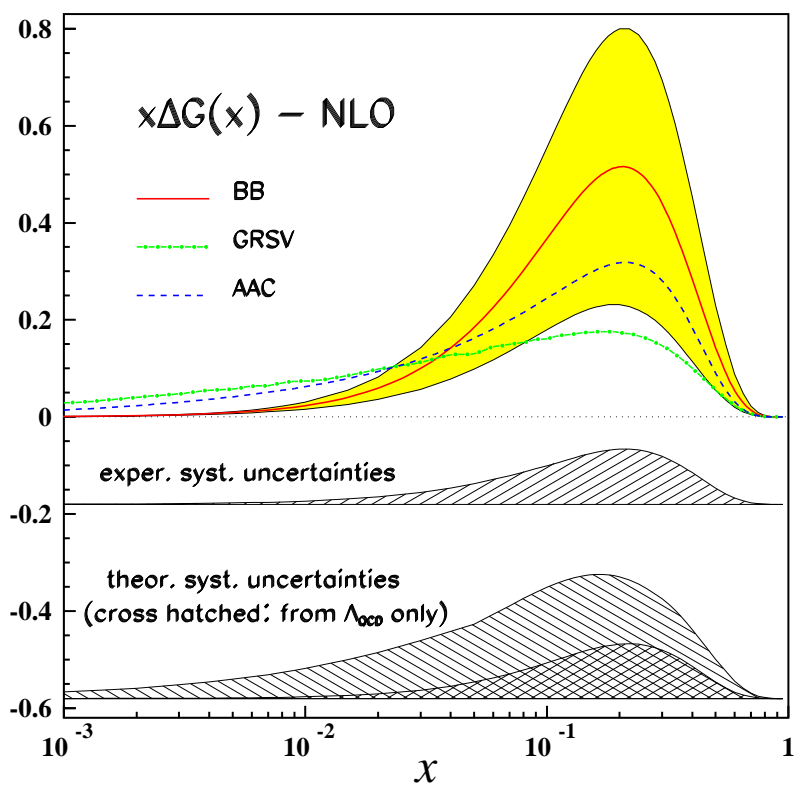}
\hspace{-2cm}
\caption{\label{pdf_comp}
\small
Polarized gluon  distribution at the input scale $Q_0^2 = 4.0~\GeV^2$
(solid line BB~\cite{a5}) compared to results
obtained by  GRSV~\cite{a4} (dashed--dotted line) and AAC~\cite{a3}
(dashed line). The shaded area represents the fully correlated
$1\sigma$ error band calculated by Gaussian error propagation.
The lower hatched areas show the present experimental systematic
uncertainty and the theory errors due to the variation of the 
renormalization and factorization scales. The pure effect of the error
in $\Lambda_{QCD}$ is also given separately\normalsize
.}
\end{center}
\end{figure}


\end{document}